\documentclass
[aps,prl,twocolumn,floatfix,english,showpacs,10pt]{revtex4-2}%
\usepackage{graphicx}
\usepackage{booktabs}
\usepackage{array}
\usepackage{makecell}
\usepackage{amsmath}
\usepackage{physics}
\usepackage{amssymb}
\usepackage{colordvi}
\usepackage{verbatim}
\usepackage{xcolor}
\usepackage{mathrsfs}
\usepackage{epsfig}
\usepackage{lipsum}
\usepackage{amsfonts}
\usepackage{makecell}
\usepackage{esint}
\usepackage{bm}
\usepackage[most]{tcolorbox}

\usepackage[unicode=true, breaklinks=false, pdfborder={0 0 1}, backref=false,
colorlinks=true, linkcolor=blue, urlcolor=blue, citecolor=blue]{hyperref}%
\providecommand{\U}[1]{\protect\rule{.1in}{.1in}}
\providecommand{\U}[1]{\protect\rule{.1in}{.1in}}
\setcitestyle{numbers,square}

\begin{document}

\title{Angular Momentum Hall Effect of Axial Ferrons in Ferroelectrics}

\author{Xiyin Ye}
\affiliation{School of Physics, Huazhong University of Science and Technology, Wuhan 430074, China}

\author{Tao Yu}
\email{taoyuphy@hust.edu.cn}
\affiliation{School of Physics, Huazhong University of Science and Technology, Wuhan 430074, China}

\date{\today }

\begin{abstract}
Phonon thermal and angular-momentum (AM) Hall effects are known in non-ferroelectric insulators, yet whether terahertz soft modes of displacive ferroelectrics, i.e., ``ferrons", share them remains open. We report axial ferrons that hold polarization-momentum locking in uniaxial ferroelectrics and show that both effects exist: a longitudinal temperature gradient drives a transverse AM flow without any field, whereas an out-of-plane magnetic field switches on the thermal Hall response. Two competitive channels emerge for the AM Hall effect: self-radiated magnetic fields hybridize two branches into circular modes whose momentum-locked Berry curvature deflects opposite magnetic moments to opposite edges, while temperature-gradient-induced mode quantum coherence adds a magnetic-moment dipole that leads to an opposite transverse AM current. An out-of-plane magnetic field breaks the exact polarization-momentum locking and yields a polarization Hall current. Numerical calculations show these effects are significant in both LiNbO$_3$ and van der Waals ferroelectrics. Edge magnetization or polarization accumulation can switch adjacent magnets or ferroelectrics, enabling low-power magnetoelectric functionality by thermal current in the future.

\end{abstract}

\maketitle

\textit{Introduction}.---Thermal Hall and angular momentum (AM) Hall effects of charge-neutral phonons~\cite{Strohm2005,Sheng2006,Inyushkin2007,Kagan2008,Zhang2010,Qin2012,Sugii2017,Grissonnanche2019,Saito2019,Boulanger2020,Grissonnanche2020,Li2020,Park2020,Chen2022,Flebus2022,Mangeolle2022,Oh2025,Lopez2026,Meng2026} make a close analogy to the ordinary Hall and spin Hall effect of electrons~\cite{Karplus1954,Nagaosa2010,vonKlitzing2020,Sinova2015,Bernevig2005,Go2018}, but are traced to distinct physical origins.   
The thermal Hall effect of phonons refers to a driven heat flow transverse to the temperature gradient under an external magnetic field, which is widely reported in insulators~\cite{Strohm2005,Inyushkin2007,Sugii2017,Grissonnanche2019,Grissonnanche2020,Boulanger2020,Li2020,Chen2022,Meng2026} and attributed to Raman-type spin-phonon interaction~\cite{Sheng2006,Kagan2008}, Berry curvature~\cite{Zhang2010,Qin2012,Saito2019}, or phonon skew scattering~\cite{Flebus2022,Mangeolle2022,Oh2025}.  The phonon AM Hall effect transports transverse AM flow and does not need a strong magnetic field. It was predicted for acoustic phonons in non-polar lattices, driven by temperature-gradient-induced mode mixing~\cite{Park2020,Lopez2026}. Here we report other dominant mechanisms for the optical phonons operating at terahertz (THz) frequencies.

The transverse optical (TO) phonon softens and induces an electric-polarization order parameter ${\bf P}$ for displacive ferroelectrics. Such soft phonons, which were recently dubbed ``ferrons" due to large static electric moments they carry~\cite{Bauer2021,Bauer2022,Tang2022,Zhuang2022,Zhou2023,Wooten2023,Bauer2023,TangBauer2024,RodriguezSuarez2024,Zhu2024,Morozovska2025}, carry dynamical electric polarizations that generate strong THz electromagnetic radiation~\cite{Choe2026,Zhang2026,Subedi2026}. Beyond this electric property, time-dependent polarization may induce a magnetic moment ${\bf P}\times \partial_t {\bf P}$~\cite{Dzyaloshinskii2009} through the circulating displacement current~\cite{Pols2026,TangBauer2026}, drawing an analogy with chiral phonons~\cite{ZhangNiu2014,ZhangNiu2015,Zhu2018,Juraschek2019,Cheng2020,Geilhufe2021,Baydin2022,Tauchert2022,Hernandez2023,Ishito2023,Basini2024,ZhangH2025} that can have a magnetic moment by dynamical multiferroicity~\cite{Juraschek2017}.  These developments naturally raise the questions of whether an analogous AM/thermal Hall effect can arise for soft phonons in ferroelectrics and, crucially, what the associated dominant mechanism is.

\begin{figure}[htp!]	
\centering
\includegraphics[width=0.47\textwidth,trim=0.0cm 1.0cm 0cm 1.3cm]{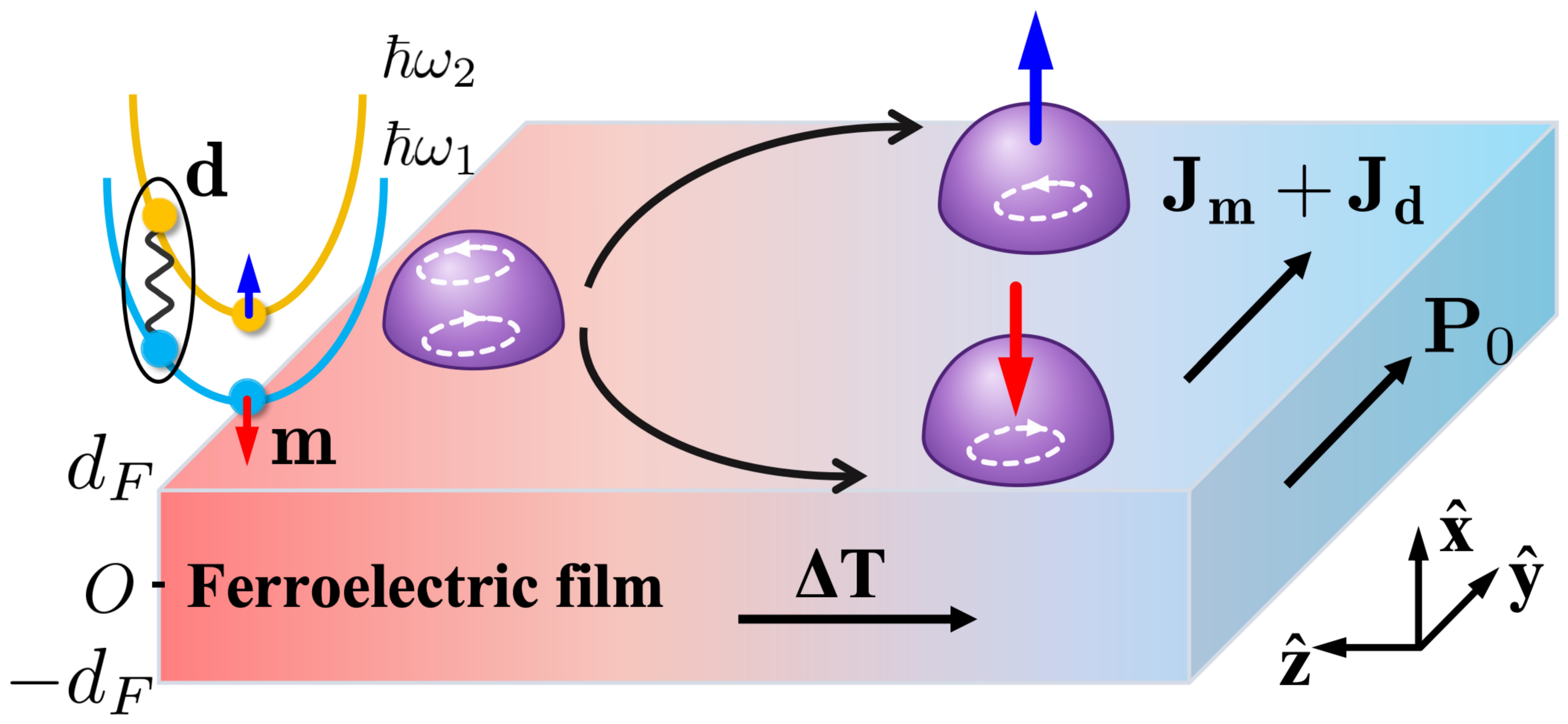}
\caption{Schematic of AM Hall effect for axial ferron wavepackets (purple) in ferroelectric films with a spontaneous polarization ${\bf P}_0\parallel \hat{\bf y}$, driven by a temperature gradient $\nabla T \parallel \hat{\bf z}$. 
Two ferron bands $\hbar\omega_1$ and $\hbar\omega_2$ carry antiparallel static magnetic moments ${\bf m}$, which are deflected by Berry curvature and generate a transverse AM current ${\bf J}_{\bf m}$. Meanwhile, the temperature gradient induces the interband coherence that endows the wavepacket with a magnetic-dipole moment ${\bf d}$, yielding a bound AM current ${\bf J}_{\bf d}$.} 
\label{model}
\end{figure}

In this Letter, we answer these questions by predicting axial ferrons and their AM Hall and thermal Hall effects in ferroelectric insulators. As illustrated in Fig.~\ref{model}, a longitudinal temperature gradient drives a transverse AM flow carried by ferron wavepackets. We find two competitive mechanisms for the ferron AM Hall effect. First, the self-radiated magnetic field hybridizes two ferron branches near the degenerate hot spots, giving oppositely oriented magnetic moments and Berry curvature at opposite momenta, with which the wavepackets carrying opposite magnetic moments are deflected to opposite directions. 
Second, the temperature gradient induces quantum coherence between different ferron modes, which contributes a magnetic-moment dipole; its rate of change leads to an opposite transverse AM current with a smaller magnitude. Further, when an out-of-plane magnetic field is applied, the Berry-curvature mechanism renders a thermal/polarization Hall current. The predicted ferron Hall effect in uniaxial ferroelectrics differs in mechanism from the simulation performed for BaTiO$_3$~\cite{Bustamante2026}; it differs from the simulated ferron-polariton Hall response as well, since the latter needs coherent THz excitation and is not linear in applied magnetic fields~\cite{Yaniv2026}. The AM/polarization accumulations at the edge by the AM/thermal Hall effect of ferrons may be used to switch the magnetization/polarization of proximity magnets/ferroelectrics.

\textit{Axial ferrons}.---We illustrate the principle by considering a ferroelectric insulator thin film of thickness $2d_F$ with surface normal along $\hat{\mathbf{x}}$ (Fig.~\ref{model}). The spontaneous electric polarization $\mathbf{P}_0=P_{0}\hat{\mathbf{y}}$ lies in the film plane. Including the electric-polarization fluctuation $\delta\mathbf{p}(\mathbf{r},t)$, the total polarization $\mathbf{P}(\mathbf{r},t)=\mathbf{P}_0+\delta\mathbf{p}(\mathbf{r},t)$. 
The fluctuation $\delta\mathbf{p}(\mathbf{r},t)$ of frequency $\omega$ radiates not only the electric $\mathbf{E}_d(\mathbf{r},t)$ but also the magnetic $\mathbf{H}_d(\mathbf{r},t)$ fields [see Supplemental Material (SM)~\cite{supplement} for details]. These fields act back on the electric fluctuation and contribute importantly to the total energy of ferroelectrics~\cite{Zhou2023,TangBauer2026,Tomeno1988,Scrymgeour2005,Hlinka2006,Chandra_book}
\begin{align}
    F&=\int_{{\bf r}\in V_f} d{\bf r}\Big(\frac{m_p}{2}|\partial_t{\bf P}|^2+f[{\bf P}]- \frac{1}{2}{\bf E}_d(\mathbf{r},t) \cdot \mathbf{P} \nonumber\\
    & -\frac{1}{2}\mu_0{\bf H}_d(\mathbf{r},t) \cdot \eta_p(\mathbf{P}\times\partial_t \mathbf{P})\Big),
    \label{free_energy}
\end{align}
which consists of kinetic and potential energies. Here, $V_f$ is the ferroelectric volume, $m_p=1/(\varepsilon _0 \Omega_p^2)$, governed by the ionic plasma frequency $\Omega_p$ and vacuum permittivity $\varepsilon_0$, is the inertial mass, and $\mu_0$ is the vacuum permeability. 
The Landau-Ginzburg-Devonshire free-energy density for uniaxial ferroelectrics $f[{\bf P}]={\alpha_1} P_y^2/2 + {\alpha_2} P_y^4/4 +{\alpha_3} (P_x^2 +P_z^2)/2$ with $\alpha_1<0$ and $\{\alpha_2,\alpha_3\}>0$; its minimization gives  $P_{0}=\sqrt{-\alpha_1/\alpha_2}$. 
In contrast to previous works~\cite{Tang2022,Zhou2023,RodriguezSuarez2024,TangBauer2024}, we here include in Eq.~(\ref{free_energy}) the Zeeman coupling between the self-radiated magnetic field ${\bf H}_d({\bf r},t)$ and the magnetization $\eta_p(\mathbf{P}\times\partial_t \mathbf{P})$ generated by the polarization current $\partial_t{\bf P}$, where $\eta_p=\gamma_p m_p$ is associated with ionic gyromagnetic ratio $\gamma_p$~\cite{Pols2026,TangBauer2026}.
Since the magnetic field ${\bf H}_d$ is generated by the polarization waves themselves, it does not break the time-reversal symmetry.

Expanding Eq.~(\ref{free_energy}) to the quadratic order of $\delta {\bf p}$ yields the Hamiltonian
\begin{align}
    \hat{H}_0&=\int_{{\bf r}\in V_f} d{\bf r} \Big[({m_p}/{2})|\partial_t \delta{\bf p}|^2+\left(1/({2\varepsilon_0})\right){\cal K}\delta {\bf p}^2 \nonumber\\
    &-(1/2){{\bf E}_d(\delta{\bf p})}\cdot \delta {\bf p}-(1/2)\mu_0\eta_p{\bf H}_d(\delta{\bf p})\cdot ({\bf P}_0\times\partial_t\delta{\bf p})\Big] \nonumber\\
    &=\hat{H}_{a}+\hat{H}_{b}+\hat{H}_{c},
\end{align}
in which, with anisotropic  $K_{\perp}=\varepsilon_0\alpha_3$ and $K_{\parallel}=\varepsilon_0(\alpha_1+3\alpha_2P_{0}^2)$, ${\cal K}=\mathrm{diag}(K_{\perp},K_{\parallel},K_{\perp})$ is a matrix of ``stiffness".
Here, $\hat H_{a}$ contains the kinetic energy,  harmonic potential, and part of the electric dipolar energies, which describes three decoupled oscillations along $\hat{\mathbf{x}}$, $\hat{\mathbf{y}}$, and $\hat{\mathbf{z}}$. The dipolar electric and magnetic fields couple these modes, which are expressed in $\hat H_{b}$ and $\hat H_{c}$, respectively.

With bosonic annihilation operator $\hat a_{\mu,\mathbf{k}}$ expressing the polarization operator 
$\delta\hat{\mathbf{p}}_{\mu,\mathbf{k}}
=\sqrt{{\hbar}/({2m_p\Omega_{\mu,\mathbf{k}}})}
(\hat a_{\mu,\mathbf{k}}+\hat a^{\dagger}_{\mu,-\mathbf{k}})$
polarized along $\mu=\{x,y,z\}$, the Hamiltonian
$\hat H_{a}=\sum_{\mu}\sum_{\mathbf{k}}\hbar\Omega_{\mu,\mathbf{k}}
(\hat a^{\dagger}_{\mu,\mathbf{k}}\hat a_{\mu,\mathbf{k}}+1/2)$ is contributed by three  harmonic oscillators with angular frequencies
\begin{align}
\Omega_{x,\mathbf{k}}&=\Omega_p\sqrt{\frac{1-e^{-2d_Fk}}{2d_Fk}+K_\perp}, \nonumber\\
\Omega_{y,\mathbf{k}}&=\Omega_p\sqrt{K_\parallel+B_kk_y^{2}},~~
\Omega_{z,\mathbf{k}}=\Omega_p\sqrt{K_\perp+B_kk_z^{2}},
\label{omega_xyz}
\end{align}
where $B_k=[2d_Fk-(1-e^{-2d_Fk})]/(2d_Fk^{3})$ and $k=\sqrt{k_y^2+k_z^2}$. The branch $\Omega_x$ is isotropic and has a higher frequency $\Omega_x\gg \{\Omega_y,\Omega_z\}$ due to the large depolarization effect $(1-e^{-2d_Fk})/2d_Fk\approx 1$. The in-plane oscillation branches $\Omega_y$ and $\Omega_z$ are strongly anisotropic; for $K_\parallel>K_\perp$, they are degenerate along the curve $K_\parallel+B_k k_y^2=K_\perp+B_k k_z^2$.

The dipolar electromagnetic field couples two in-plane fluctuations in distinct ways: the electric
dipolar field produces a real coupling constant
$g_{\mathbf{k}}={\hbar}
{B_kk_yk_z}/({2m_p\varepsilon_0}\sqrt{\Omega_{y,\mathbf{k}}\Omega_{z,\mathbf{k}}})$,
while the magnetic dipolar field acting on the
oscillating magnetic moment $\mathbf{m}^a=\eta_p\mathbf{P}_0\times\partial_t\delta\mathbf{p}$ produces an imaginary one
\begin{equation}
i\chi_{\mathbf{k}}=(i{\hbar}/{4})\gamma_p\mu_0P_{0}B_kk_z
\sqrt{\Omega_{y,\mathbf{k}}\Omega_{z,\mathbf{k}}},
\end{equation}
such that 
\begin{align}
    \hat{H}_{b}&=\frac{1}{2}\sum_{\bf k} g_{\bf k}\left(\hat{a}^{\dagger}_{y,{\bf k}}+\hat{a}_{y,-{\bf k}}\right)\left(\hat{a}_{z,{\bf k}}+\hat{a}^{\dagger}_{z,-{\bf k}}\right)+{\rm H.c.}, \nonumber\\
    \hat{H}_{c}&=-\frac{i}{2}\sum_{\bf k}{\chi}_{\bf k}\left(\hat{a}^{\dagger}_{y,{\bf k}}-\hat{a}_{y,-{\bf k}}\right)\left(\hat{a}_{z,{\bf k}}-\hat{a}^{\dagger}_{z,-{\bf k}}\right)+{\rm H.c.}.
    \nonumber
\end{align}
The out-of-plane fluctuation $\delta p_x$ does not mix with the other modes. Focusing on the fluctuations along the $\hat{\bf y}$- and $\hat{\bf z}$-directions, we collect them into the four-component spinor $\hat{\Psi}_{\bf k}=(\hat{a}_{y,{\bf k}},\hat{a}_{z,{\bf k}},\hat{a}^{\dagger}_{y,-{\bf k}},\hat{a}^{\dagger}_{z,-{\bf k}})^T$ and  write the Hamiltonian $\hat{H}_{0}\rightarrow (1/2)\sum_{\bf k}\hat{\Psi}_{\bf k}^{\dagger}{\cal H}_{0}({\bf k})\hat{\Psi}_{\bf k}$ in terms of the Bogoliubov-de~Gennes Hamiltonian matrix 
\begin{align}
    {\cal H}_{0}({\bf k})=\begin{pmatrix}
        \hbar\Omega_{y,{\bf k}} & \kappa_{\bf k} & 0 & \kappa^{\ast}_{\bf k}\\
        \kappa^{\ast}_{\bf k} & \hbar\Omega_{z,{\bf k}} & \kappa_{\bf k} & 0 \\
        0 &  \kappa^{\ast}_{\bf k} & \hbar\Omega_{y,-{\bf k}} & \kappa_{\bf k} \\
        \kappa_{\bf k} & 0 & \kappa^{\ast}_{\bf k} & \hbar\Omega_{z,-{\bf k}}    \label{H_effk}   
        \end{pmatrix},
\end{align}
in which $\kappa_{\bf k}=g_{\bf k}-i\chi_{\bf k}$. Crucially, the real $g_{\bf k}$ mixes the amplitudes of $\delta p_y$ and $\delta p_z$ and causes level repulsion, whereas the imaginary $i\chi_{\bf k}$ mixes their phases and is the source of circular polarization.

\begin{figure}[htp!]	
\centering
\includegraphics[width=0.48\textwidth,trim=0.0cm 0.45cm 0cm 0.45cm]{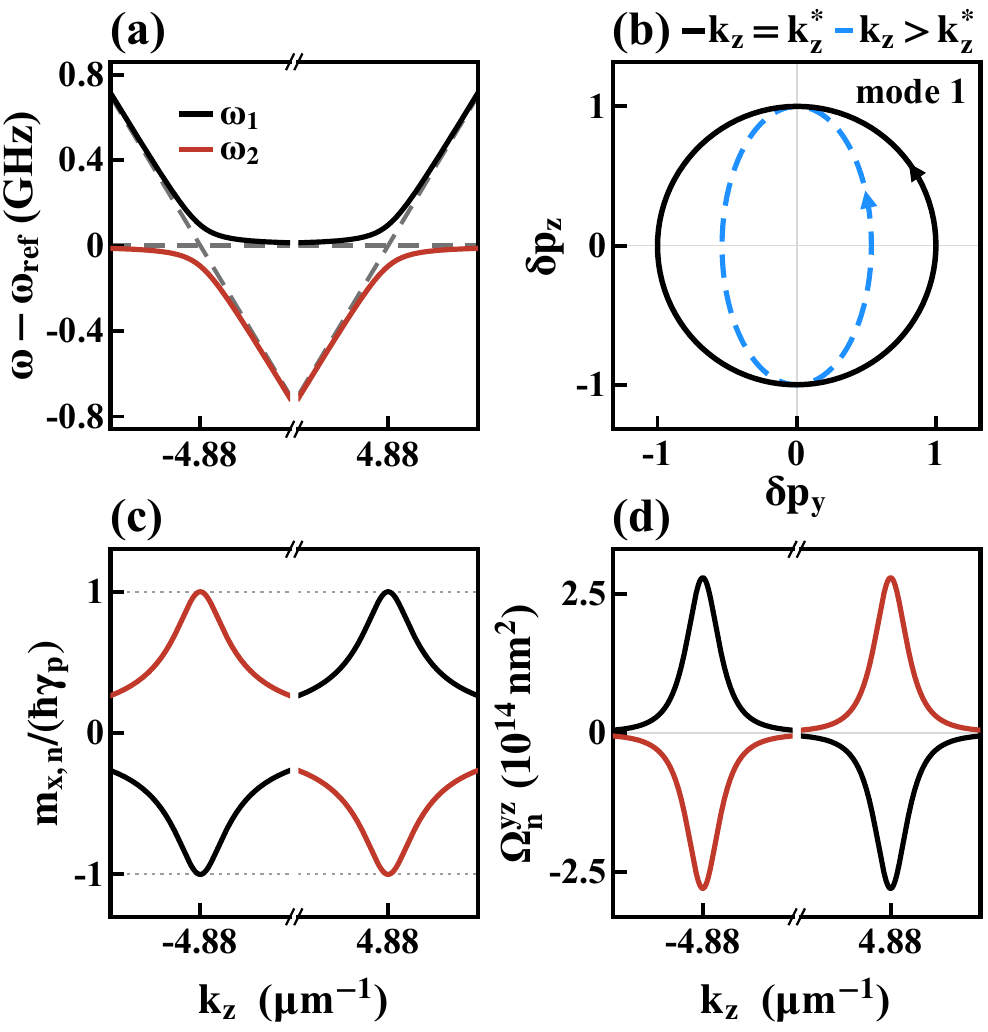}
\caption{(a) plots the dispersions of two ferron modes along $k_z$, measured from $\omega_{\rm ref}= \Omega_{y}(k_y=0)=47.4$~THz. (b) shows the rotation of in-plane polarization fluctuations for mode 1 near the anticrossing point. (c) and (d): Static magnetic moment and Berry curvature of two branches. All panels are evaluated with $k_y=0$.} 
\label{curvature}
\end{figure}

The degeneracy condition $K_\parallel+B_k k_y^2=K_\perp+B_k k_z^2$ of Eq.~(\ref{omega_xyz}) has many solutions. However, since \(g_{\mathbf{k}}\propto k_y k_z\), all degeneracies with \(k_y\neq0\) are lifted. At \(k_y=0\), \(g_{\mathbf{k}}=0\) while \(\chi_{\mathbf{k}}\propto k_z\) remains finite. Consequently, the avoided crossing at $k_z=\pm k_z^{\ast}$, determined by \(K_\parallel=K_\perp+B_{k_z^{\ast}}k_z^{\ast 2}\), is opened solely by the magnetic back-action through a purely phase-like mixing, and its gap $2|\chi_{\mathbf{k}}|$ directly measures the magnetoelectric coupling.
To obtain the associated ferron modes, we diagonalize Eq.~(\ref{H_effk}) via the Bogoliubov transformation $\hat\Psi_{\mathbf{k}}=T_{\mathbf{k}}\hat X_{\mathbf{k}}$, where
$\hat X_{\mathbf{k}}=(\hat b_{1,\mathbf{k}},\hat b_{2,\mathbf{k}},\hat b^{\dagger}_{1,-\mathbf{k}},\hat b^{\dagger}_{2,-\mathbf{k}})^{T}$ is composed of ferron operators $\hat b_{1,\mathbf{k}}$ and $\hat b_{2,\mathbf{k}}$. The $4 \times 4$ transformation matrix $T_{\mathbf{k}}$ is paraunitary in that  $T_{\mathbf{k}}\sigma_3T^{\dagger}_{\mathbf{k}}=\sigma_3$ with 
$\sigma_3=\mathrm{diag}(1,1,-1,-1)$. This yields $\hat H_0=(1/2)\sum_{\mathbf{k}}\hat X^{\dagger}_{\mathbf{k}}E_{\mathbf{k}}\hat X_{\mathbf{k}}$, where $E_{\bf k}={\rm diag}(\hbar\omega_{1,{\bf k}},\hbar\omega_{2,{\bf k}},\hbar\omega_{1,-{\bf k}},\hbar\omega_{2,-{\bf k}})$ is the ferron dispersion. Figure~\ref{curvature}(a) plots the ferron frequency spectra for LiNbO$_3$ film with thickness $2d_F=10$~nm, relative to $\omega_{\rm ref}\equiv \Omega_{y}(k_y=0)=47.4$~THz, showing an avoided crossing between two branches $\hbar\omega_{1,\mathbf{k}}$ and $\hbar\omega_{2,\mathbf{k}}$ at $k_y=0$ and $k_z=\pm k_z^{\ast}$. For LiNbO$_3$, the ionic plasma frequency $\Omega_p=250$~THz, $\{\alpha_1,\alpha_3\}=\{-2.012,1.345\}\times10^{9}$~${\rm N}\cdot {\rm m}^2/{\rm C}^2$, and $\alpha_2=3.608\times10^{9}$~${\rm N}\cdot {\rm m}^6/{\rm C}^4$, yielding dimensionless stiffness $\{K_{\perp},K_{\parallel}\}=\{0.012,0.036\}$ and equilibrium polarization $P_0=0.746$~C/m$^2$ at room temperature~\cite{Scrymgeour2005,Zhou2023}. Van der Waals ferroelectric NbOI$_2$ shows similar feature with $\Omega_p=78.4$~THz, $\{K_{\perp},K_{\parallel}\}=\{0.06,0.063\}$, and $P_0=0.2$~C/m$^2$~\cite{Choe2026,wu2022,subedi2025}.
Throughout, we take $\gamma_p=10^{-2}\gamma_e$, where $\gamma_e$ is the gyromagnetic ratio of electrons.

The hybridization near this avoided crossing leads to circular polarizations. To see this, we retain only the particle-conserving $\hat{a}_{\bf k}^\dagger \hat{a}_{\bf k}$ terms in Eq.~(\ref{H_effk}) since the terms involving two-particle creation/annihilation give only small squeezing to circular polarization (refer to a complete calculation in the SM~\cite{supplement}). The in-plane motion of $\delta {\bf p}$ then reduces to an effective two-level problem. 
At $k_y=0$ and $k_z=\pm k_z^{\ast}$, the amplitudes have a fixed ratio ${\delta p_{z,n}(t)}/{\delta p_{y,n}(t)}=s_n\,e^{i\phi_{\mathbf{k}}}$,
where the phase angle $\phi_{\mathbf{k}}$ is implied via $\kappa_{\mathbf{k}}=|\kappa_{\mathbf{k}}|e^{-i\phi_{\mathbf{k}}}$, and $s_{1,2}=\pm $ for two ferron modes $n=\{1,2\}$. 
Since $\kappa_{\mathbf{k}}=-i\chi_{\mathbf{k}}$ at $k_y=0$,  one has
$\phi_{\mathbf{k}}=\mathrm{sgn}(k_z)\pi/2$. For $k_z=k_z^*>0$, the ratio is $\pm i$ for $n=\{1,2\}$, viz., mode 1(2) is right(left)-circularly polarized.
Away from $k_z^{\ast}$, the finite splitting between $\Omega_{y,\mathbf{k}}$ and
$\Omega_{z,\mathbf{k}}$ makes oscillations along $\hat{\bf y}$ and $\hat{\bf z}$ unequal in amplitude, rendering an elliptical polarization while preserving the opposite
handedness of the two modes. 
Figure~\ref{curvature}(b) illustrates the polarization of mode 1. Reversing the propagation reverses $\phi_{\mathbf{k}}$, such that for modes of the same frequency the handedness is reversed. This polarization-momentum locking is the ferron analogue of the valley-dependent circular dichroism of gapped Dirac systems~\cite{Xiao2012,Cao2012,Mak2012,Zeng2012,Xu2014,Yu2014} and valley-locked chirality of phonons~\cite{Zhu2018}. An applied out-of-plane magnetic field breaks the time-reversal symmetry as well as the polarization-momentum locking. Strictly speaking, the polarization axis $\hat{\bf L}$ of these circular motions is along the surface normal $\hat{\bf x}$-direction, which is thereby normal to the ferron propagation direction, so these modes are ``axial achiral" with vanished helicities $\hat{\bf L}\cdot {\bf k}=0$~\cite{Chiral_phonons}.

For ferrons, the handedness is not merely a polarization property. A circulating electric polarization constitutes a loop electric current and therefore carries a magnetic moment. Moreover, \(\phi_{\mathbf{k}}\) sets the wave-vector dependent phase between \(\delta p_y\) and \(\delta p_z\) in the underlying ferron mode. At the same time, the Berry curvature measures exactly such \(\mathbf{k}\)-dependence in the mixed polarization. Both quantities are addressed as follows.

\textit{Magnetic moment versus Berry curvature}.---The magnetic moment density generated by the circulating current is ${\mathbf{m}}_{\rm tot}(\mathbf{r})=\eta_p\,\mathbf{P}\times\partial_t\mathbf{P}
={\bf m}^a(\mathbf{r})+{\bf m}^b(\mathbf{r})$.
Two contributions are distinguishable.
The first term,
$\mathbf{m}^a({\bf r})=\eta_p\mathbf{P}_0\times\partial_t\delta\mathbf{p}$, oscillates at the ferron frequency and yields no static magnetic moment.
However, it gives rise to $\chi_{\mathbf{k}}$ via coupling to $\mathbf{H}_d$, which renders the modes handed.
The static magnetic moment comes entirely from the second term ${\bf m}^b({\bf r})=\eta_p\,\delta{\mathbf{p}}\times\partial_t\delta{\mathbf{p}}$, which exists only for circularly or elliptically polarized $\delta\mathbf{p}$.
Because $\delta\mathbf{p}$ rotates in the $yz$ plane, ${\bf m}^b({\bf r})$ aligns along $\hat{\bf x}$, normal to both $\mathbf{P}_0$ and the propagation direction, so ferrons here couple directly to an out-of-plane external magnetic field.
The total magnetic moment $\hat{M}_{x}=\int \hat m^b_{x}({\bf r}) d{\bf r}=(1/2)\sum_{\bf k}\hat{X}_{\bf k}^{\dagger}\tilde{\cal M}_{{\bf k}}\hat{X}_{\bf k}$, in which $\tilde{\cal M}_{{\bf k}}=T^{\dagger}_{\bf k}{\cal M}_{{\bf k}}T_{\bf k}$. With $\mu_{\pm,{\bf k}}=\hbar\gamma_p(\sqrt{\Omega_{z,{\bf k}}/\Omega_{y,{\bf k}}}\pm\sqrt{\Omega_{y,{\bf k}}/\Omega_{z,{\bf k}}})/2$, the matrix
\begin{align}
    {\cal M}_{{\bf k}}=\begin{pmatrix}
         0 & -i\mu_{+,\bf k} & 0 & i\mu_{-,\bf k}\\
        i\mu_{+,\bf k}& 0 & i\mu_{-,\bf k} & 0 \\
        0 &  -i\mu_{-,\bf k} & 0 & i\mu_{+,\bf k} \\
        -i\mu_{-,\bf k} & 0 & -i\mu_{+,\bf k} & 0   
        \end{pmatrix}.
        \label{Meq}
    \end{align}
The diagonal term $\tilde{\cal M}_{{\bf k}}|_{nn}\equiv m_{x,n}({\bf k})$ represents the static magnetic moment carried by the $n$-th ferron mode, while the off-diagonal term $\tilde{\cal M}_{{\bf k}}|_{n\ne n'}$ is contributed by the quantum coherence of different modes.

Figure~\ref{curvature}(c) shows $m_{x,n}(k_z)$ at different $k_z$ along $k_y=0$. Near the anticrossing point ($k_z\approx k_z^*$), the two modes carry opposite magnetic moments, i.e., $m_{x,1}\simeq+\hbar\gamma_p$ and $m_{x,2}\simeq-\hbar\gamma_p$. Away from the anticrossing point, $|m_{x,n}|$ decreases since the polarization becomes elliptical. Reversing the propagation reverses the sense of rotation and hence the sign of $m_{x,n}$, consistent with the time-reversal operation over a magnetic moment. The ferron magnetic moment is therefore band- and momentum-resolved, with its sign locked to the handedness.

The ${\bf k}$-dependent relative phase $\phi_{\bf k}$ between $\delta p_y$ and $\delta p_z$ is the Berry connection, and its curl gives the Berry curvature in the $\sigma_3$-metric. For right eigenvectors $\sigma_3{\cal H}_{0}({\bf k})|u_{n,\mathbf{k}} \rangle=\varepsilon_{n,{\bf k}}|u_{n,\mathbf{k}}\rangle$, with $\varepsilon_{n,{\bf k}}=(\sigma_3E_{\bf k})_{nn}$, the Berry curvature reads
\[
\Omega^{ij}_{n,\mathbf{k}}=-2\hbar^2\sum_{m\neq n}(\sigma_3)_{nn}(\sigma_3)_{mm}
\frac{\mathrm{Im}\big[(v_{{\bf k},i})_{nm}(v_{{\bf k},j})_{mn}\big]}
{(\varepsilon_{n,\mathbf{k}}-\varepsilon_{m,\mathbf{k}})^{2}},
\]
in which $(v_{{\bf k},i})_{nm}=(1/\hbar)\langle u_{n,\mathbf{k}}|\partial_{k_i}{\cal H}_0(\mathbf{k})|u_{m,\mathbf{k}}
\rangle$. The Berry curvature $\Omega_{n,{\bf k}}^{ij}(-\chi_{\bf k})=-\Omega_{n,{\bf k}}^{ij}(\chi_{\bf k})$ vanishes when $\chi_{\bf k}\rightarrow 0$. Figure~\ref{curvature}(d) shows $\Omega^{yz}_{n,\mathbf{k}}(k_z)$ at $k_y=0$. As an odd function of ${\bf k}$, i.e., $\Omega_{n}(-\mathbf{k})=-\Omega_{n}(\mathbf{k})$, the product $m_{x,n}(\mathbf{k})\Omega_{n,\mathbf{k}}$ is even in $\mathbf{k}$, leading to one mechanism of AM Hall response in ferroelectrics.

\textit{Ferron AM Hall effect}.---We recognize two competitive mechanisms responsible for the ferron AM Hall effect. The magnetic moment carried by ferrons is not conserved since $[\hat{M}_x,\hat{H}_0]\neq 0$~\cite{Xiao2021,Koyama2025}. The torque $\hat{\tau}^{M_x}({\bf r})=(i/2\hbar)(\hat{H}_0\sigma_3\hat{m}^b_x({\bf r})-\hat{m}^b_x({\bf r})\sigma_3\hat{H}_0)$ does not vanish, and the conventional AM current $\hat{j}^{M_x}_{\alpha}({\bf r})=(1/4)(\hat{\bf v}\sigma_3\hat{m}^b_x({\bf r})+\hat{m}^b_x({\bf r})\sigma_3\hat{\bf v})$ alone does not define transport. Based on
$\langle\hat{\tau}^{M_x}\rangle=-\nabla\cdot\boldsymbol{\pi}^{M_x}$, we define the conserved AM current (see SM for its calculation~\cite{supplement})
\begin{equation}
J^{L_x}_{\alpha}
=\frac{1}{\gamma_p V_f}\int d\mathbf{r}\,
\Big(\left\langle\hat{j}^{M_x}_{\alpha}\right\rangle+\pi^{M_x}_{\alpha}\Big).
\end{equation}
These quantities are evaluated in terms of the ferron wavepacket $|W_n\rangle$ in branch $n$, which is localized at 
$\mathbf{r}_c\equiv\langle W_n|\sigma_3\hat{\mathbf{r}}|W_n\rangle
 /\langle W_n|\sigma_3|W_n\rangle$
and $\mathbf{k}$~\cite{wave_packet,RMP,Xiao2021,Koyama2025}. 
The temperature gradient enters via the local Bose-Einstein distribution $f_B(T(\mathbf{r}),\varepsilon)$, which drives transverse AM flow $J^{L_x}_{\alpha}=\kappa_{\alpha\beta}\nabla_{\beta}T$ of ferron via the conductivity tensor~\cite{supplement}
\begin{align}
\kappa_{\alpha\beta}
&=-\frac{k_B}{\hbar\gamma_pV_f}\sum_{n=1}^{2}\sum_{\bf k}
\left[\Omega^{\alpha\beta}_{n,\mathbf{k}}m_{x,n}
-\frac{\partial}{\partial k_{\beta}}\left(d^{M_x}_{n,\mathbf{k}}\right)_\alpha\right]\nonumber\\
&\times c_1\left[f_B(\varepsilon_{n,{\bf k}})\right],
\label{kappaAM}
\end{align}
with $c_1[f]=(1+f)\ln(1+f)-f\ln f$ and  $f_B(\varepsilon)=(e^{\beta \varepsilon}-1)^{-1}$ with $\beta=1/(k_BT_0)$. Here, the torque-induced magnetic-moment dipole  reads
\begin{align}
    &\left(d^{M_x}_{n,\mathbf{k}}\right)_{\alpha}={\rm Re}~\big\langle W_{n} \big| (\hat{r}_{\alpha}-r_{c,\alpha})\hat{M}_x\big| W_{n} \big\rangle /\langle W_n|\sigma_3|W_n\rangle \nonumber\\
    &=\hbar\sum^4_{n'(\neq n)=1}(\sigma_{3})_{nn}(\sigma_{3})_{n'n'}
\frac{\mathrm{Im}\left[(v_{\mathbf{k},\alpha})_{nn'}\big(\tilde{\cal M}_{x,\mathbf{k}}\big)_{n'n}\right]}{{\varepsilon}_{\mathbf{k},n}-{\varepsilon}_{\mathbf{k},n'}}.
    \nonumber
\end{align}
Thereby, the transport of magnetic moment is enabled not only by the band-diagonal $m_{x,n}$ carried by ferron modes, which is directly deflected by the Berry curvature, but also by an interband quantum coherence $(\tilde{\cal M}_{x,\mathbf{k}})_{n'n}$ induced between different modes.

The second term in Eq.~\eqref{kappaAM} corresponds to a bound AM current from the magnetic-moment dipole $\big(d^{M_x}_{n,\mathbf{k}}\big)_{\alpha}$ carried by ferron wavepackets, analogous to the bound electric current ${\bf J}=\partial_t {\bf P}$ arising from the electric polarization. Indeed, a localized wavepacket carrying no net AM can still possess AM dipole moments $D^{M_x}_{\alpha}=(1/V_f)\sum^2_{n=1}\sum_{\bf k}\big(d^{M_x}_{n,\mathbf{k}}\big)_{\alpha}f_B(T({\bf r}),\varepsilon_{n,{\bf k}})$; its rate of change generates a bound AM current
\begin{align}
    &J^{L_x}_{D,\alpha}=\frac{1}{\gamma_p}\partial_t D_\alpha=\frac{1}{\gamma_pV_f}\sum^2_{n=1}\sum_{\bf k}\big(d^{M_x}_{n,\mathbf{k}}\big)_{\alpha}\partial_tf_B(\varepsilon_{n,{\bf k}},T({\bf r}))\nonumber\\
    &\left. \approx\frac{1}{\gamma_pV_f}\sum^2_{n=1}\sum_{\bf k}\big(d^{M_x}_{n,\mathbf{k}}\big)_{\alpha}
    (v_{{\bf k},\beta})_{nn}\frac{\partial f_B(\varepsilon_{n,{\bf k}},T({\bf r}))}{\partial T}\right|_{T_0}\nabla_{\beta}T\nonumber\\
    &=\frac{k_B}{\hbar\gamma_pV_f}\sum^2_{n=1}\sum_{\bf k}\left[\partial_{k_{\beta}} \big(d^{M_x}_{n,\mathbf{k}}\big)_{\alpha}\right]c_1\left[f_B(\varepsilon_{n,{\bf k}},T_0)\right]\nabla_{\beta}T,
    \nonumber
\end{align}
which is exactly the second term in Eq.~\eqref{kappaAM}. Thereby, even when ferron modes carry no static magnetic moment $m_{x,n}$, as for linearly polarized ferrons, the interband coherence $(\tilde{\cal M}_{x,\mathbf{k}})_{n'\ne n}$ can still yield AM Hall responses.

\begin{figure}[htp!]	
\centering
\includegraphics[width=0.48\textwidth,trim=0.0cm 1.1cm 0cm 0.6cm]{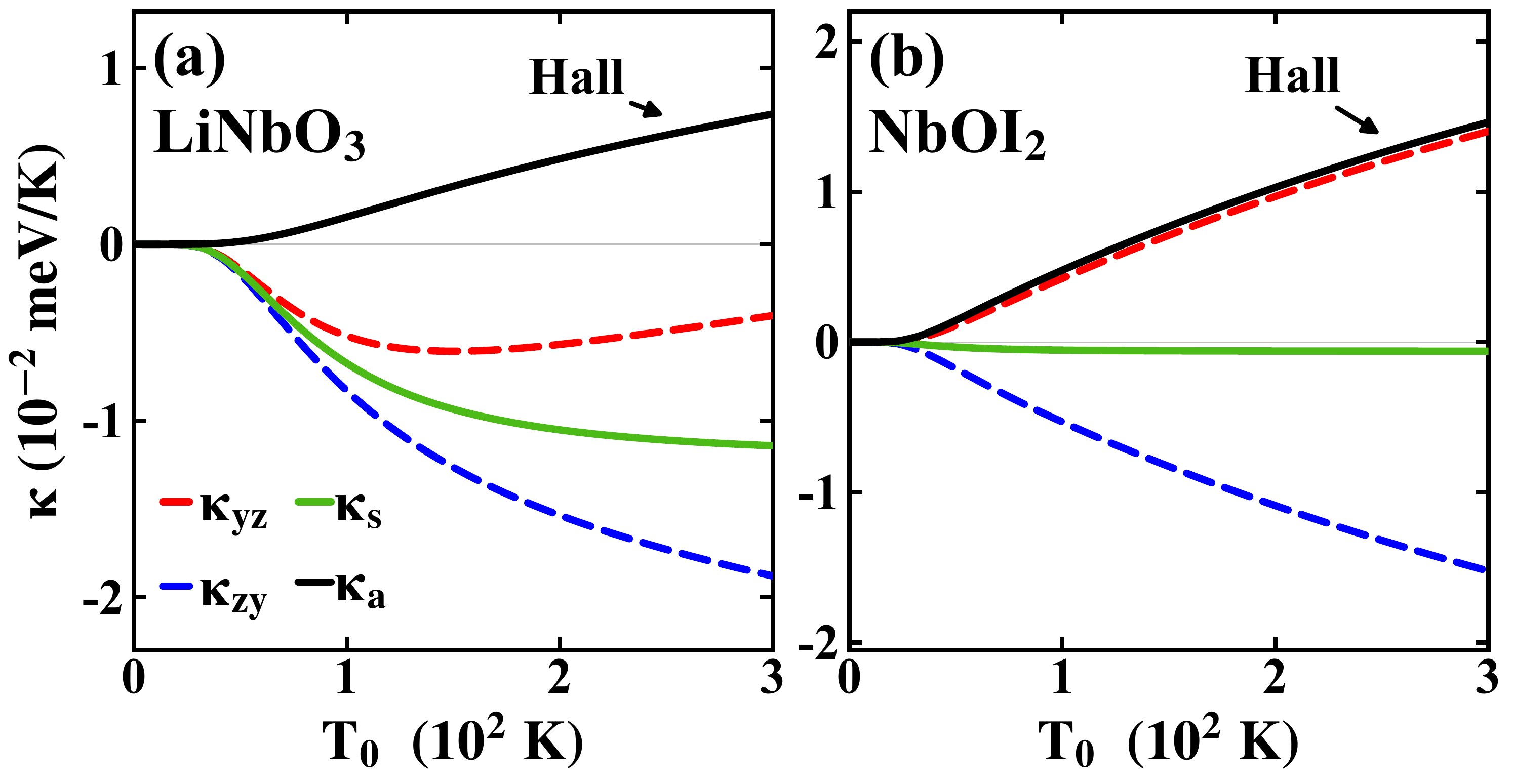}
\caption{AM Hall conductivities versus temperature for LiNbO$_3$ [(a)] and NbOI$_2$ [(b)]. $\kappa_s$ and $\kappa_a$ denote the symmetric and antisymmetric combinations of $\kappa_{yz}$ and $\kappa_{zy}$.} 
\label{kappaT}
\end{figure}

In the conductivity, $\kappa_a=(\kappa_{yz}-\kappa_{zy})/2$ is the genuine antisymmetric Hall response contributed by both terms. The comparison in End Matter shows the Berry curvature effect dominates over the magnetic-moment dipole. $\kappa_s=(\kappa_{yz}+\kappa_{zy})$ is solely carried by the magnetic-moment dipole (see End Matter).  Figure~\ref{kappaT}(a) compares $\kappa_a$ and $\kappa_s$ for LiNbO$_3$ as a function of temperature. $\kappa_{yz}$ and $\kappa_{zy}$ differ but have the same sign, leading to a significant genuine Hall response $\kappa_a$ and a large $\kappa_s$ as well at $300$~K, with $|\kappa_a/\kappa_s|\approx 0.57$. Figure~\ref{kappaT}(b) shows the opposite situation for NbOI$_2$: $\kappa_s$ almost vanishes while $\kappa_a$ dominates, giving $|\kappa_a/\kappa_s|\approx 22$ at $300$~K.
All components decay exponentially below $30$~K, at which the thermal population of ferrons near the anticrossing hot spot is depleted. 

\textit{Ferron thermal Hall effect}.---An out-of-plane external magnetic field ${\bf B}\parallel\hat{\bf x}$ breaks the time-reversal symmetry through Zeeman coupling with the magnetic moment Eq.~(\ref{Meq}), thereby turning on the thermal Hall response of ferrons while leaving the AM Hall effect essentially unchanged. The thermal Hall conductivity originates from the Berry curvature, which reads~\cite{Matsumoto2011,Matsumoto2014,Qin2012,Chern2021}
\begin{align}
    \sigma_{\alpha\beta}=-\frac{k_B^2T_0}{\hbar V_f}\sum^2_{n=1}\sum_{\bf k}\Omega^{\alpha\beta}_{n,\mathbf{k}}c_2\left[f_B(\varepsilon_{n,{\bf k}})\right],
\end{align}
where $c_2(\rho)=(1+\rho)(\ln [(1+\rho)/\rho])^2-(\ln \rho)^2-2{\rm Li_{2}}(-\rho)$ with ${\rm Li_{2}}(z)$ being the polylogarithm function.
On the other hand, the longitudinal thermal conductivity, governed by the relaxation time $\tau$, reads
\begin{align}
    \sigma_{\alpha\alpha}=\frac{\tau}{k_BT^2_0 V_f}\sum^2_{n=1}\sum_{\bf k}(v_{\mathbf{k},\alpha})^2_{nn}\varepsilon^2_{n,{\bf k}}\frac{e^{\beta\varepsilon_{n,{\bf k}}}}{(e^{ \beta\varepsilon_{n,{\bf k}}}-1)^2}.
\end{align}
We estimate the Hall angle $\tan\theta\equiv\sigma_{yz}/\sigma_{zz}$ by choosing $B=30$~T, $T_0=300$~K, and $\tau=1$~ps. For NbOI$_2$, $\sigma_{yz}/(2d_F)\approx 0.13~{\rm mW ~K^{-1}~m^{-1}}$ and $\sigma_{zz}/(2d_F)\approx 61.4~{\rm mW ~K^{-1}~m^{-1}}$ lead to $\tan \theta \approx 2\times 10^{-3}$, which should be detectable by recent experiments~\cite{Meng2026}.

\textit{Conclusion and discussion}.---In summary, we show that a THz soft mode in uniaxial ferroelectrics deflects its own angular momentum under a temperature gradient. 
Its handedness is self-generated: the electric polarization radiates the magnetic field that mixes two modes, so no magnetic field is needed to break time-reversal symmetry. Two contributions, i.e., Berry-curvature deflection and coherence-enabled bound current of magnetic-moment dipole, compete; their relative weight is material-dependent and experimentally accessible. An out-of-plane magnetic field then acts as a tuning handle, not a prerequisite, converting the AM flow into a polarization Hall current. Different from the phonon Hall effect in non-polar materials that borrows chirality from phonon-spin coupling, skew scattering, or mode mixing by applied bias, the soft phonon supplies it via its own radiated field, a criterion that extends to many polar modes with dynamical electric polarization, including multiaxial ferroelectrics, tunable van der Waals layers, and ferroelectric-magnet heterostructures.

\begin{acknowledgments}
This work is financially supported by the National Natural Science Foundation of China under Grant No.~12374109 and the National Key Research and Development Program of China under Grant No.~2023YFA1406600. 
\end{acknowledgments}

\section{End Matter}

\begin{figure}[htp!]	
\centering
\includegraphics[width=0.48\textwidth,trim=0.0cm 1.1cm 0cm 0.4cm]{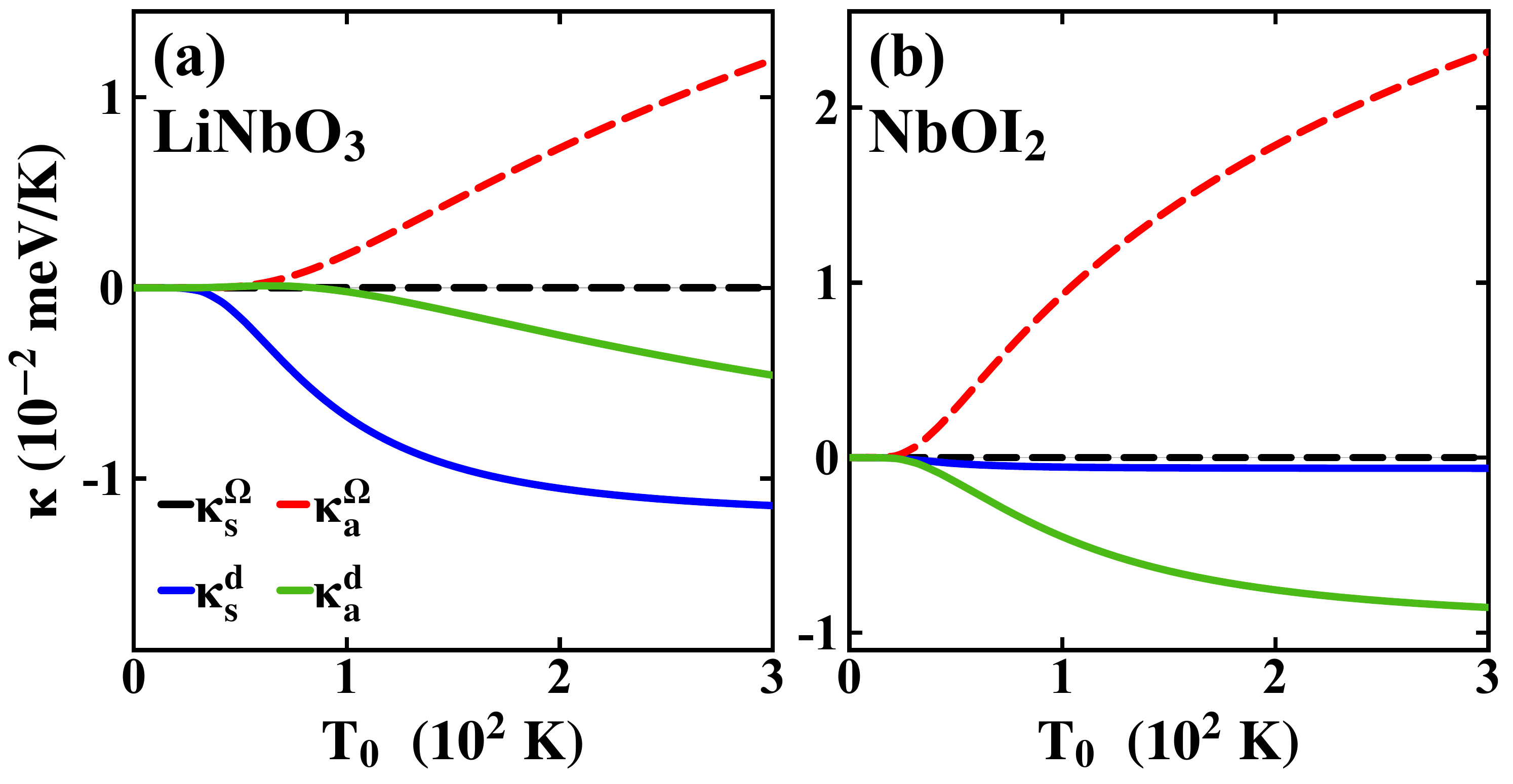}
\caption{Berry-curvature versus magnetic-moment--dipole contributions to the ferron angular-momentum Hall conductivity of LiNbO$_3$ [(a)] and NbOI$_2$ [(b)], decomposed into symmetric $\kappa_s^{\Omega/d}$ and antisymmetric $\kappa_a^{\Omega/d}$ components.} 
\label{kappa_components}
\end{figure}

\textit{Comparison of two mechanisms in AM Hall effect}.---Here, we compare in detail the contributions of Berry curvature and magnetic-moment dipole to the AM Hall effect (in the absence of $B\hat{\bf x}$).
Figure~\ref{kappa_components} distinguishes the two mechanisms by comparing their contributions to the ferron AM Hall conductivity of LiNbO$_3$ and NbOI$_2$ at different temperatures. For the symmetric component $\kappa_s$, the contribution from Berry curvature $\kappa_s^\Omega$ vanishes, so it arises entirely from the dipole contribution $\kappa_s^d$. This is because the Berry curvature $\Omega_{yz}=-\Omega_{zy}$ is antisymmetric. By contrast, for the antisymmetric component $\kappa_a$, the contribution from the magnetic-moment dipole $\kappa_a^d$ partially cancels that from the Berry curvature $\kappa_a^{\Omega}$ in both materials.

\begin{thebibliography}{99}

\bibitem{Strohm2005} C. Strohm, G. L. J. A. Rikken, and P. Wyder, Phenomenological evidence for the phonon Hall effect, Phys. Rev. Lett. \textbf{95}, 155901 (2005).

\bibitem{Sheng2006} L. Sheng, D. N. Sheng, and C. S. Ting, Theory of the phonon Hall effect in paramagnetic dielectrics, Phys. Rev. Lett. \textbf{96}, 155901 (2006).

\bibitem{Inyushkin2007} A. V. Inyushkin and A. N. Taldenkov, On the phonon Hall effect in a paramagnetic dielectric, JETP Lett. \textbf{86}, 379 (2007).

\bibitem{Kagan2008} Yu. Kagan and L. A. Maksimov, Anomalous Hall effect for the phonon heat conductivity in paramagnetic dielectrics, Phys. Rev. Lett. \textbf{100}, 145902 (2008).

\bibitem{Zhang2010} L. Zhang, J. Ren, J.-S. Wang, and B. Li, Topological nature of the phonon Hall effect, Phys. Rev. Lett. \textbf{105}, 225901 (2010).

\bibitem{Qin2012} T. Qin, J. Zhou, and J. Shi, Berry curvature and the phonon Hall effect, Phys. Rev. B \textbf{86}, 104305 (2012).

\bibitem{Sugii2017} K. Sugii, M. Shimozawa, D. Watanabe, Y. Suzuki, M. Halim, M. Kimata, Y. Matsumoto, S. Nakatsuji, and M. Yamashita, Thermal Hall effect in a Phonon-Glass Ba$_3$CuSb$_2$O$_9$, Phys. Rev. Lett. \textbf{118}, 145902 (2017).

\bibitem{Grissonnanche2019} G. Grissonnanche, A. Legros, S. Badoux, E. Lefran\c{c}ois, V. Zatko, M. Lizaire, F. Lalibert\'e, A. Gourgout, J.-S. Zhou, S. Pyon, \textit{et al.}, Giant thermal Hall conductivity in the pseudogap phase of cuprate superconductors, Nature \textbf{571}, 376 (2019).

\bibitem{Saito2019} T. Saito, K. Misaki, H. Ishizuka, and N. Nagaosa, Berry phase of phonons and thermal Hall effect in nonmagnetic insulators, Phys. Rev. Lett. \textbf{123}, 255901 (2019).

\bibitem{Boulanger2020} M.-E. Boulanger, G. Grissonnanche, S. Badoux, A. Allaire, \'E. Lefran\c{c}ois, A. Legros, A. Gourgout, M. Dion, C. H. Wang, X. H. Chen, \textit{et al.}, Thermal Hall conductivity in the cuprate Mott insulators Nd$_2$CuO$_4$ and Sr$_2$CuO$_2$Cl$_2$, Nat. Commun. \textbf{11}, 5325 (2020).

\bibitem{Grissonnanche2020} G. Grissonnanche, S. Th\'eriault, A. Gourgout, M.-E. Boulanger, E. Lefran\c{c}ois, A. Ataei, F. Lalibert\'e, M. Dion, J.-S. Zhou, S. Pyon, \textit{et al.}, Chiral phonons in the pseudogap phase of cuprates, Nat. Phys. \textbf{16}, 1108 (2020).

\bibitem{Li2020} X. Li, B. Fauqu\'e, Z. Zhu, and K. Behnia, Phonon thermal Hall effect in strontium titanate, Phys. Rev. Lett. \textbf{124}, 105901 (2020).

\bibitem{Park2020} S. Park and B.-J. Yang, Phonon angular momentum Hall effect, Nano Lett. \textbf{20}, 7694 (2020).

\bibitem{Chen2022} L. Chen, M.-E. Boulanger, Z.-C. Wang, F. Tafti, and L. Taillefer, Large phonon thermal Hall conductivity in the antiferromagnetic insulator Cu$_3$TeO$_6$, Proc. Natl. Acad. Sci. U.S.A. \textbf{119}, e2208016119 (2022).

\bibitem{Flebus2022} B. Flebus and A. H. MacDonald, Charged defects and phonon Hall effects in ionic crystals, Phys. Rev. B \textbf{105}, L220301 (2022).

\bibitem{Mangeolle2022} L. Mangeolle, L. Balents, and L. Savary, Phonon thermal Hall conductivity from scattering with collective fluctuations, Phys. Rev. X \textbf{12}, 041031 (2022).

\bibitem{Oh2025} T. Oh and N. Nagaosa, Phonon thermal Hall effect in Mott insulators via skew scattering by the scalar spin chirality, Phys. Rev. X \textbf{15}, 011036 (2025).

\bibitem{Lopez2026} D. A. Bustamante Lopez, V. Brehm, and D. M. Juraschek, Atomistic theory of the phonon angular momentum Hall effect, arXiv:2604.01899 (2026).

\bibitem{Meng2026} Q. Meng, X. Li, J. Liu, L. Zhao, C. Dong, Z. Zhu, L. Li, and K. Behnia, Thermodynamic origin of the phonon Hall effect in a honeycomb antiferromagnet, Science \textbf{394}, 419 (2026).



\bibitem{Karplus1954} R. Karplus and J. M. Luttinger, Hall effect in ferromagnetics, Phys. Rev. \textbf{95}, 1154 (1954). 

\bibitem{Nagaosa2010} N. Nagaosa, J. Sinova, S. Onoda, A. H. MacDonald, and N. P. Ong, Anomalous Hall effect, Rev. Mod. Phys. \textbf{82}, 1539 (2010).

\bibitem{vonKlitzing2020} K. von Klitzing, T. Chakraborty, P. Kim, V. Madhavan, X. Dai, J. McIver, Y. Tokura, L. Savary, D. Smirnova, A. M. Rey, C. Felser, J. Gooth, and X. Qi, 40 years of the quantum Hall effect, Nat. Rev. Phys. \textbf{2}, 397 (2020).


\bibitem{Sinova2015} J. Sinova, S. O. Valenzuela, J. Wunderlich, C. H. Back, and T. Jungwirth, Spin Hall effects, Rev. Mod. Phys. \textbf{87}, 1213 (2015). 

\bibitem{Bernevig2005} B. A. Bernevig, T. L. Hughes, and S.-C. Zhang, Orbitronics: The intrinsic orbital current in $p$-doped silicon, Phys. Rev. Lett. \textbf{95}, 066601 (2005). 

\bibitem{Go2018} D. Go, D. Jo, C. Kim, and H.-W. Lee, Intrinsic spin and orbital Hall effects from orbital texture, Phys. Rev. Lett. \textbf{121}, 086602 (2018).






 

\bibitem{Choe2026} J. Choe, T. Handa, C.-Y. Huang, A. K. Liston, J. Cox, J. Stensberg, Y. Hong, D. G. Chica, D. Xu, F. Tay, \textit{et al.}, Observation of coherent ferron emission and propagation, Nat. Mater. \textbf{25}, 1755 (2026).

\bibitem{Zhang2026} B. Zhang, R. Duan, S. S. Mishra, S. Jana, J. Kim,
T. T. Caiwei, Y. J. Tan, W. Wang, P. T. C. Ietro, Z. Liu, and R. Singh,
Electric-field control of coherent ferron oscillations,
Nat. Phys. {\bf 22}, 1624 (2026).

\bibitem{Subedi2026} S. Subedi, W. Fang, F. Fei, Z. Zhai, J. P. Rollins, C. Fox, A. Drew, B. Lv, Y. Ping, and J. Xiao,
Electrically switchable ferron upconversion in a van der Waals ferroelectric, arXiv:2603.19394.




\bibitem{Bauer2021} G. E. W. Bauer, R. Iguchi, and K.-i. Uchida, Theory of transport in ferroelectric capacitors, Phys. Rev. Lett. \textbf{126}, 187603 (2021). 

\bibitem{Bauer2022} G. E. W. Bauer, P. Tang, R. Iguchi, and K.-i. Uchida, Magnonics vs. ferronics, J. Magn. Magn. Mater. \textbf{541}, 168468 (2022). 





\bibitem{Tang2022} P. Tang, R. Iguchi, K.-i. Uchida, and G. E. W. Bauer, Excitations of the ferroelectric order, Phys. Rev. B \textbf{106}, L081105 (2022). 

\bibitem{Zhuang2022} S. Zhuang and J.-M. Hu, Role of polarization-photon coupling in ultrafast terahertz excitation of ferroelectrics, Phys. Rev. B \textbf{106}, L140302 (2022). 

\bibitem{Zhou2023} X.-H. Zhou, C. Cai, P. Tang, R. L. Rodr\'iguez-Su\'arez, S. M. Rezende, G. E. W. Bauer, and T. Yu, Surface ferron excitations in ferroelectrics and their directional routing, Chin. Phys. Lett. \textbf{40}, 087103 (2023).

\bibitem{Wooten2023} B. L. Wooten, R. Iguchi, P. Tang, J. S. Kang, K.-i. Uchida, G. E. W. Bauer, and J. P. Heremans, Electric field-dependent phonon spectrum and heat conduction in ferroelectrics, Sci. Adv. \textbf{9}, eadd7194 (2023).

\bibitem{Bauer2023} G. E. W. Bauer, P. Tang, R. Iguchi, J. Xiao, K. Shen, Z. Zhong, T. Yu, S. M. Rezende, J. P. Heremans, and K.-i. Uchida, Polarization transport in ferroelectrics, Phys. Rev. Appl. \textbf{20}, 050501 (2023).

\bibitem{TangBauer2024} P. Tang and G. E. W. Bauer, Electric analog of magnons in order-disorder ferroelectrics, Phys. Rev. B \textbf{109}, L060301 (2024). 

\bibitem{RodriguezSuarez2024} R. L. Rodr\'iguez-Su\'arez, X.-H. Zhou, C. Y. Cai, P. Tang, T. Yu, G. E. W. Bauer, and S. M. Rezende, Surface and volume modes of polarization waves in ferroelectric films, Phys. Rev. B \textbf{109}, 134307 (2024). 

\bibitem{Zhu2024} Y. Zhu, T. Chen, A. Ross, B. Wang, X. Guo, V. Gopalan, L.-Q. Chen, and J.-M. Hu, Theory of nonlinear terahertz susceptibility in ferroelectrics, Phys. Rev. B \textbf{110}, 054311 (2024). 

\bibitem{Morozovska2025} A. N. Morozovska, E. A. Eliseev, O. V. Bereznikov, M. Ye. Yelisieiev, G.-D. Zhao, Y. Zhu, V. Gopalan, L.-Q. Chen, J.-M. Hu, and Y. M. Vysochanskii, Flexocoupling-induced phonons and ferrons in van der Waals ferroelectrics, Phys. Rev. B \textbf{112}, 014110 (2025). 



\bibitem{Dzyaloshinskii2009} I. Dzyaloshinskii and D. Mills, Intrinsic paramagnetism of ferroelectrics, Philos. Mag. \textbf{89}, 2079 (2009). 

\bibitem{Pols2026} M. Pols, C. P. Romao, and D. M. Juraschek, Lattice Excitations with Finite Polarization and Magnetization, Phys. Rev. Lett. \textbf{136}, 166801 (2026). 


\bibitem{TangBauer2026} P. Tang and G. E. W. Bauer, Multiferroic-like Quasiparticles in Ferroelectrics, Phys. Rev. Lett. \textbf{136}, 176702 (2026). 



\bibitem{ZhangNiu2014} L. Zhang and Q. Niu, Angular momentum of phonons and the Einstein--de Haas effect, Phys. Rev. Lett. \textbf{112}, 085503 (2014). 

\bibitem{ZhangNiu2015} L. Zhang and Q. Niu, Chiral phonons at high-symmetry points in monolayer hexagonal lattices, Phys. Rev. Lett. \textbf{115}, 115502 (2015). 

\bibitem{Zhu2018} H. Zhu, J. Yi, M.-Y. Li, J. Xiao, L. Zhang, C.-W. Yang, R. A. Kaindl, L.-J. Li, Y. Wang, and X. Zhang, Observation of chiral phonons, Science \textbf{359}, 579 (2018). 

\bibitem{Juraschek2019} D. M. Juraschek and N. A. Spaldin, Orbital magnetic moments of phonons, Phys. Rev. Mater. \textbf{3}, 064405 (2019). 

\bibitem{Cheng2020} B. Cheng, T. Schumann, Y. Wang, X. Zhang, D. Barbalas, S. Stemmer, and N. P. Armitage, A large effective phonon magnetic moment in a Dirac semimetal, Nano Lett. \textbf{20}, 5991 (2020). 

\bibitem{Geilhufe2021} R. M. Geilhufe, V. Juri\v{c}i\'c, S. Bonetti, J.-X. Zhu, and A. V. Balatsky, Dynamically induced magnetism in KTaO$_3$, Phys. Rev. Res. \textbf{3}, L022011 (2021). 

\bibitem{Baydin2022} A. Baydin, F. G. G. Hernandez, M. Rodriguez-Vega, A. K. Okazaki, F. Tay, G. T. Noe, I. Katayama, J. Takeda, H. Nojiri, P. H. O. Rappl, E. Abramof, G. A. Fiete, and J. Kono, Magnetic control of soft chiral phonons in PbTe, Phys. Rev. Lett. \textbf{128}, 075901 (2022). 

\bibitem{Tauchert2022} S. R. Tauchert, M. Volkov, D. Ehberger, D. Kazenwadel, M. Evers, H. Lange, A. Donges, A. Book, W. Kreuzpaintner, U. Nowak, \textit{et al.}, Polarized phonons carry angular momentum in ultrafast demagnetization, Nature \textbf{602}, 73 (2022). 

\bibitem{Hernandez2023} F. G. G. Hernandez, A. Baydin, S. Chaudhary, F. Tay, I. Katayama, J. Takeda, H. Nojiri, A. K. Okazaki, P. H. O. Rappl, E. Abramof, M. Rodriguez-Vega, G. A. Fiete, and J. Kono, Observation of interplay between phonon chirality and electronic band topology, Sci. Adv. \textbf{9}, eadj4074 (2023). 


\bibitem{Ishito2023} K. Ishito, H. Mao, Y. Kousaka, Y. Togawa, S. Iwasaki, T. Zhang, S. Murakami, J.-i. Kishine, and T. Satoh, Truly chiral phonons in $\alpha$-HgS, Nat. Phys. \textbf{19}, 35 (2023). 

\bibitem{Basini2024} M. Basini, M. Pancaldi, B. Wehinger, M. Udina, V. Unikandanunni, T. Tadano, M. C. Hoffmann, A. V. Balatsky, and S. Bonetti, Terahertz electric-field-driven dynamical multiferroicity in SrTiO$_3$, Nature \textbf{628}, 534 (2024). 


\bibitem{ZhangH2025} H. Zhang, N. Peshcherenko, F. Yang, T. Z. Ward, P. Raghuvanshi, L. Lindsay, C. Felser, Y. Zhang, J.-Q. Yan, and H. Miao, Measurement of phonon angular momentum, Nat. Phys. \textbf{21}, 1387 (2025). 



\bibitem{Juraschek2017} D. M. Juraschek, M. Fechner, A. V. Balatsky, and N. A. Spaldin, Dynamical multiferroicity, Phys. Rev. Mater. \textbf{1}, 014401 (2017).



\bibitem{Bustamante2026} D. A. Bustamante Lopez, V. Brehm, and D. M. Juraschek, Ferron Hall effect: Transverse accumulation of polarization driven by thermal gradients in ferroelectrics, arXiv:2606.29765 (2026). 

\bibitem{Yaniv2026} O. Yaniv and D. M. Juraschek, Ferron polariton Hall effect: Deflecting optically rectified polarization with magnetic fields, arXiv:2611.05025 (2026). 


\bibitem{supplement} See Supplementary Material [...] for the detailed derivation of the radiated electric and magnetic fields by fluctuation of electric polarization, the calculation of the Berry curvature of ferrons, and the formulation of angular-momentum thermal Hall effect for ferrons in terms of wave-packet formalism.




\bibitem{Tomeno1988} I. Tomeno and S. Matsumura, Dielectric properties of LiTaO$_3$, Phys. Rev. B \textbf{38}, 606 (1988).


\bibitem{Scrymgeour2005} D. A. Scrymgeour, V. Gopalan, A. Itagi, A. Saxena, and P. J. Swart, Phenomenological theory of a single domain wall in uniaxial trigonal ferroelectrics: Lithium niobate and lithium tantalate, Phys. Rev. B \textbf{71}, 184110 (2005).


\bibitem{Hlinka2006} J. Hlinka and P. M{\'a}rton, Phenomenological model of a $90^{\circ}$ domain wall in BaTiO$_3$-type ferroelectrics, Phys. Rev. B \textbf{74}, 104104 (2006).

\bibitem{Chandra_book} P. Chandra and P. B. Littlewood, \textit{A Landau Primer for Ferroelectrics in Physics of Ferroelectrics}, pp 69–116, (Springer, Berlin, 2007).


\bibitem{wu2022} Y. Wu, I. Abdelwahab, K. C. Kwon, I. Verzhbitskiy,
L. Wang, W. H. Liew, K. Yao, G. Eda, K. P. Loh, L. Shen, and S. Y. Quek,
Data-driven discovery of high performance layered van der Waals
piezoelectric NbOI$_2$, Nat. Commun. \textbf{13}, 1884 (2022).

\bibitem{subedi2025} S. Subedi, W. Liu, W. Fang, C. Fox, Z. Zhai, F. Fei,
Y. Ping, B. Lv, and J. Xiao, Colossal terahertz emission with ultrafast
tunability based on van der Waals ferroelectric NbOI$_2$,
Adv. Opt. Mater. \textbf{13}, 2403471 (2025).


\bibitem{Xiao2012} D. Xiao, G.-B. Liu, W. Feng, X. Xu, and W. Yao, Coupled spin and valley physics in monolayers of MoS$_2$ and other group-VI dichalcogenides, Phys. Rev. Lett. \textbf{108}, 196802 (2012).

\bibitem{Cao2012} T. Cao, G. Wang, W. Han, H. Ye, C. Zhu, J. Shi, Q. Niu, P. Tan, E. Wang, B. Liu, and J. Feng, Valley-selective circular dichroism of monolayer molybdenum disulphide, Nat. Commun. \textbf{3}, 887 (2012).

\bibitem{Mak2012} K. F. Mak, K. He, J. Shan, and T. F. Heinz, Control of valley polarization in monolayer MoS$_2$ by optical helicity, Nat. Nanotechnol. \textbf{7}, 494 (2012).

\bibitem{Zeng2012} H. Zeng, J. Dai, W. Yao, D. Xiao, and X. Cui, Valley polarization in MoS$_2$ monolayers by optical pumping, Nat. Nanotechnol. \textbf{7}, 490 (2012).

\bibitem{Xu2014} X. Xu, W. Yao, D. Xiao, and T. F. Heinz, Spin and pseudospins in layered transition metal dichalcogenides, Nat. Phys. \textbf{10}, 343 (2014).

\bibitem{Yu2014} T. Yu and M. W. Wu, Valley depolarization due to intervalley and intravalley electron-hole exchange interactions in monolayer MoS$_2$, Phys. Rev. B \textbf{89}, 205303 (2014).

\bibitem{Chiral_phonons} D. M. Juraschek, R. M. Geilhufe, H. Zhu, M. Basini, P. Baum, A. Baydin, S. Chaudhary, M. Fechner, B. Flebus, G. Grissonnanche \textit{et al.}, Chiral phonons, Nat. Phys. \textbf{21}, 1532 (2025).



\bibitem{Xiao2021} C. Xiao and Q. Niu, Conserved current of nonconserved quantities, Phys. Rev. B \textbf{104}, L241411 (2021).

\bibitem{Koyama2025} S. Koyama and J. Nasu, Formulation of the spin Nernst effect for spin-nonconserving insulating magnets, Phys. Rev. B \textbf{112}, 014447 (2025).

\bibitem{wave_packet} G. Sundaram and Q. Niu, Wave-packet dynamics in slowly perturbed crystals: Gradient corrections and Berry-phase effects, Phys. Rev. B \textbf{59}, 14915 (1999).

\bibitem{RMP} D. Xiao, M.-C. Chang, and Q. Niu, Berry phase effects on electronic properties, Rev. Mod. Phys. \textbf{82}, 1959 (2010).




\bibitem{Matsumoto2011} R. Matsumoto and S. Murakami, Rotational motion of magnons and the thermal Hall effect, Phys. Rev. B \textbf{84}, 184406 (2011).



\bibitem{Matsumoto2014} R. Matsumoto, R. Shindou, and S. Murakami, Thermal Hall effect of magnons in magnets with dipolar interaction, Phys. Rev. B \textbf{89}, 054420 (2014).

\bibitem{Chern2021} L. E. Chern, E. Z. Zhang, and Y. B. Kim, Sign structure of thermal Hall conductivity and topological magnons for in-plane field polarized Kitaev magnets, Phys. Rev. Lett. \textbf{126}, 147201 (2021).



\end{thebibliography}
\end{document}